\documentclass{elsart1p}
\usepackage{graphicx}
\usepackage{amsmath}
\usepackage{amssymb}
\begin{document}
\begin{frontmatter}

\title{Statistical Thermal Models in High-Energy Nuclear Physics}

\author{Ludwik Turko}
\ead{turko@ift.uni.wroc.pl}
\address{Institute of Theoretical Physics, University of Wroc{\l}aw\\
pl. Maksa Borna 9, 50-204 Wroc{\l}aw, Poland}
\thanks{This work was supported in part by the Polish Ministry of Science and  Higher Education under contract No. N N202 0953 33.}
\begin{abstract}
An examination of thermal models leads to the important signature of the expected critical behavior of the hadronic matter. A presentation is mainly devoted to the final volume effects. Canonical suppression factor are calculated.
\end{abstract}
\end{frontmatter}
%
\section{Introduction}
\label{intro}
High-energy nuclear physics, as encountered in heavy ion collisions, is a good working example for the fruitful using of statistical physics concepts \cite{rev:2004}. Thermal models, originating from Koppe, Fermi, and Hagedorn \cite{KFH} ideas, became a strong tool to explain a lot of observed global properties of highly excited hadronic matter. Within the statistical approach, particle production can only be described using the grand canonical or canonical ensemble with respect to conservation laws, if the number of produced particles that carry a conserved charge is sufficiently large. These are statistical ensembles of non--interacting particles with decay channels taken into accounts.

There are, however, some peculiarities, which should be taken into account, to make models realistic ones. These are such as internal symmetries, both abelian and nonabelian ones and careful discrimination between canonical and grand canonical distributions. This last property is due to the effects of final hadronization volume in the considered systems.

An examination of thermal models leads to the important signature of the expected critical behavior of the hadronic matter. Charge fluctuations in the event by event analysis of heavy ion collisions can be such an effect.
A presentation is mainly devoted to the final volume effects and to the analysis of the fluctuations in critical regions of the excited hadronic matter. These problems are of crucial importance for the differences between the grand canonical and canonical description of excited hadronic matter.

\section{Statement of the problem}

Let us consider\footnote{for technical details see \cite{crt:2005}} the statistical model of a non--interacting  gas constrained by the conservation of the abelian charge $Q$. The thermodynamic system of volume $V$ and temperature $T$ is considered to be
composed of charged particles and their antiparticles carrying charge $\pm 1$
respectively. The requirement of charge conservation in the system is imposed on the
grand canonical (GC) or canonical (C)  level.

The partition functions of  the above canonical and grand canonical
statistical system have a form
  \begin{equation}\label{part funct}
\mathcal{Z}_Q^{C}(V,T)=
\sum\limits_{N_+-N_-=Q}^\infty\frac{z^{N_-+N_+}}{N_-!N_+!} = I_Q(2Vz_0)\,;\quad
\mathcal{Z}^{GC}(V,T) = \exp{\left(2Vz_0\cosh{\frac{\mu}{T}}\right)}\,.
\end{equation}
where $Vz_0$ is the sum over all  one-particle partition functions

The chemical potential $\mu$ determines the average charge in the grand canonical ensemble
\[\langle Q\rangle=T\frac{\partial}{\partial\mu}\ln\mathcal{Z}^{GC}\,.\]%
This allows to eliminate the chemical potential from further formulae for the grand
canonical probabilities distributions
\begin{equation*}\label{chem potQ}
    \frac{\mu}{T}=\text{arcsinh}\frac{\langle Q\rangle}{2 V z_0} =
    \ln\frac{\langle Q\rangle + \sqrt{\langle Q\rangle^2+4 (Vz_0)^2}}{2 V
    z_0}\,.
\end{equation*}

The probability distribution $\mathcal{P}_Q^{C}(N_-,V)$ to have $N_-$
negatively  and $N_+=N_-+Q$ positively charged particles is obtained
from the partition function  as
 \begin{equation*}\label{prob C N}
    \mathcal{P}_Q^{C}(N_-,V)= \frac{(Vz_0)^{2N_-+Q}}{N_-!(N_-+Q)!}\frac{1}{I_Q(2 Vz_0)}\,.
\end{equation*}

  In the GC ensemble with volume $V$ and  average charge $\langle
Q\rangle $ the probability distribution to have $N_-$ negative particles is
\begin{equation*}\label{probab GC neg n}
    \mathcal{P}_{\langle Q\rangle}^{GC}(N_-,V) =
    \frac{1}{N_-!}\left[\frac{2(Vz_0)^2}{\langle Q\rangle + \sqrt{\langle Q\rangle^2+4
    (Vz_0)^2}}\right]^{N_-}\,
    \exp\left[-\frac{2(Vz_0)^2}{\langle Q\rangle + \sqrt{\langle Q\rangle^2+4
    (Vz_0)^2}}\right]
\end{equation*}

  The thermodynamic limit is understood as a limit $V\to\infty$ such that densities of
the system remain constant. This corresponds to

\[Q,N_-\to\infty;\ \frac{Q}{V}=q\,;\ \frac{N_-}{V}=n_-\,,\qquad\langle Q\rangle,N_-\to\infty;\ \frac{\langle Q\rangle}{V}=\langle q\rangle\,;\
 \frac{N_-}{V}=n_-\,,\]%
for the canonical ensemble and for the grand canonical ensemble respectively.

To formulate correctly  the thermodynamic limit of quantities involving densities,
one defines  probabilities for densities
  \begin{equation}
{\mathbf{P}}_{q}^{C}(n_-,V):=  V\mathcal{P}_{V q}^{C}(V
n_-,V)\,,\qquad
{\mathbf{P}}_{\langle  q\rangle}^{GC}(q,V):=  V\mathcal{P}_{V\langle
q\rangle}^{GC}(V q,V)\,,\label{prob GC dens}
 \end{equation}
for the canonical and grand canonical ensemble respectively.

This allows \cite{crt:2005} to calculate density moments for both statistical distributions with finite volume corrections taken into account. One gets
\begin{subequations}\label{TC moments_NLO}
\begin{equation}
  \langle n_-^k\rangle^C\ \simeq\ \langle n_-\rangle_\infty^k
      - \frac{k}{V}\frac{q+\langle n_-\rangle_{\infty}}{(q+2\langle n_-\rangle_{\infty})^2}\,\langle n_-\rangle_\infty^k +
    \frac{k(k-1)}{2V}\frac{q+\langle n_-\rangle_{\infty}}{q+2\langle n_-\rangle_{\infty}}\,\langle
    n_-\rangle_\infty^{k-1}\,,
\end{equation}
for the canonical ensemble and
\begin{equation}
      \langle {n_-^k}\rangle^{GC}\simeq\langle n_-\rangle_{\infty}^k +
      \frac{k(k-1)}{2V}\,\langle n_-\rangle_{\infty}^{k-1}\,.
\end{equation}
the grand canonical ensemble.
\end{subequations}

$\langle n_-\rangle_{\infty}$ is an average limiting  density of negatively charged particles
\begin{equation}\label{dens part pm}
    \langle n_-\rangle_{\infty}= \left.\frac{\sqrt{q^2+4z_0^2} - q}{2}\right|_{q=\langle q\rangle}
\end{equation}

\begin{figure}[!th]
\includegraphics[width=0.45\textwidth]{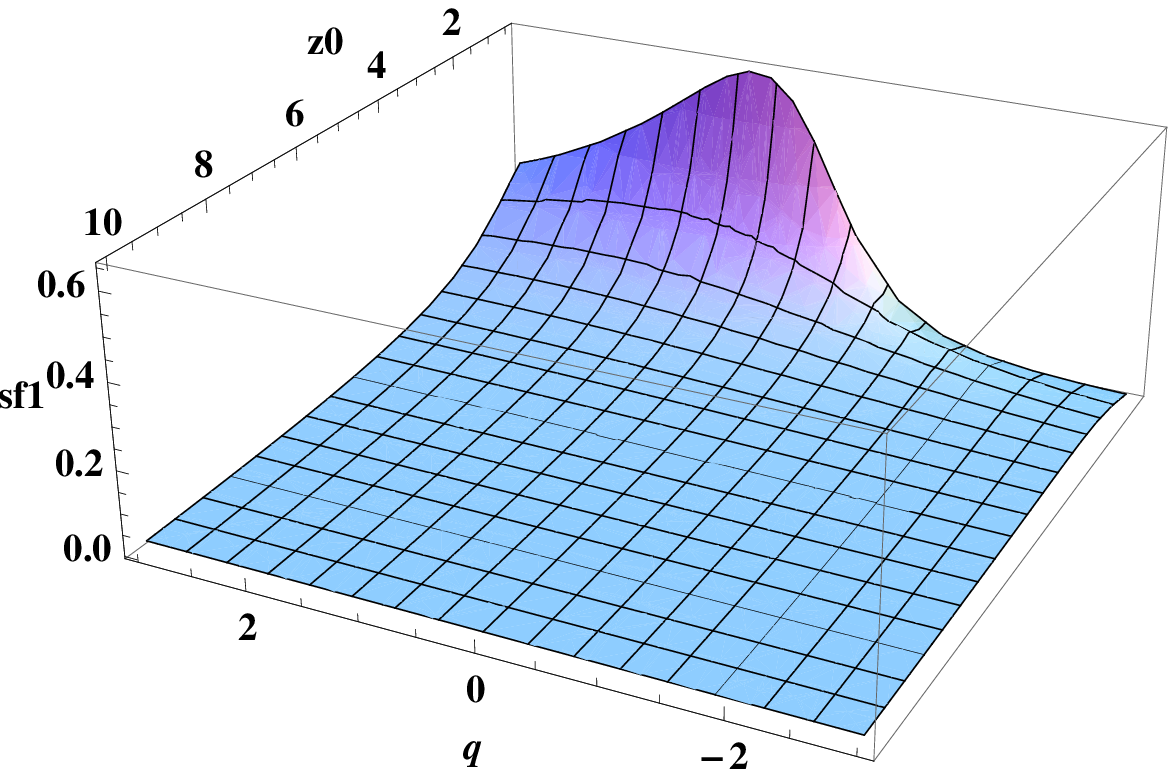}\qquad
\includegraphics[width=0.5\textwidth]{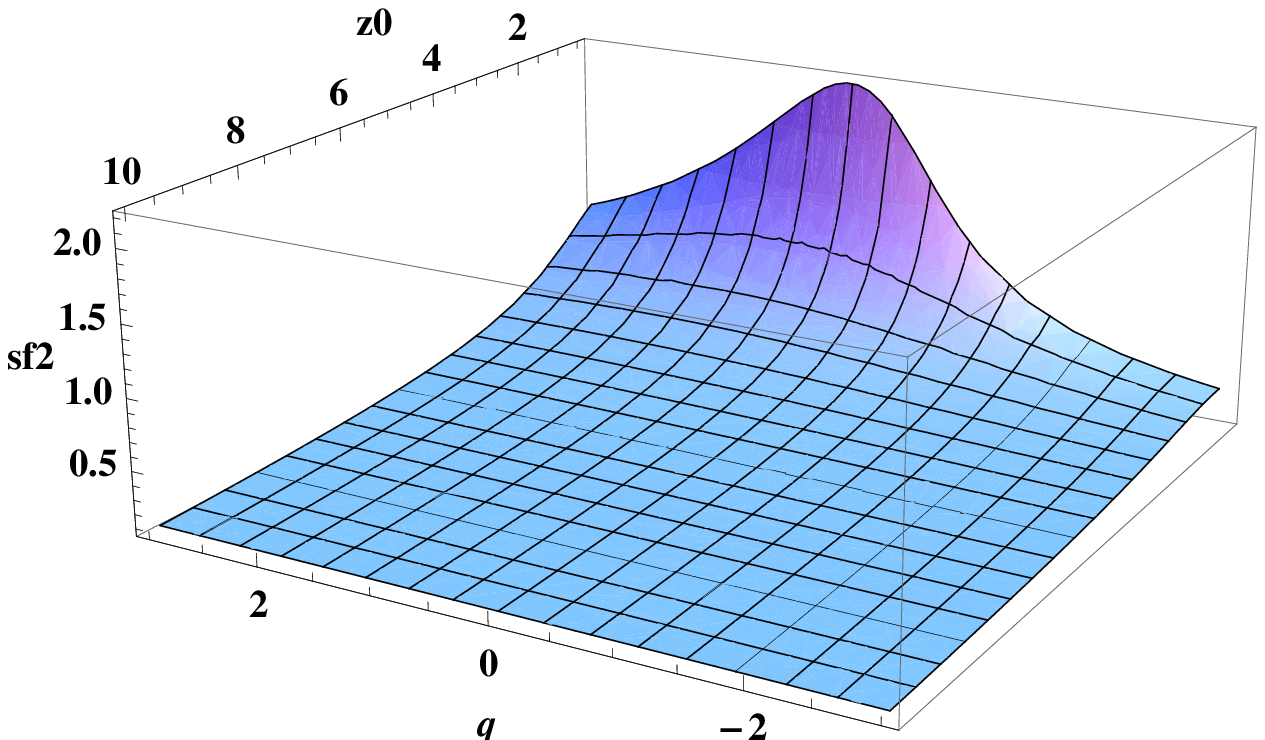}
\caption{\label{Fig.1}$1/V$ corrections to canonical suppression factors for the first and the second moments}
\end{figure}

Eqs \eqref{TC moments_NLO} allow to write the canonical suppression factor for densities as
  \begin{equation}
    \frac{\langle n_-^k\rangle_q^C}{\langle {n_-^k}\rangle^{GC}_{\langle q\rangle}} \simeq 1 -
  \frac{1}{V}\,\frac{k\langle q\rangle + k^2\sqrt{\langle q\rangle^2 +4 z_0^2}}{2(\langle q\rangle^2 +4 z_0^2)}\,.
  \end{equation}
A sketchy behavior of the first two moments is given on the Fig. \eqref{Fig.1}. It is seen that volume corrections became more and more important for higher order moments and they are decreasing for more charged systems.

\end{document}